\documentclass[lettersize,journal]{IEEEtran}
\IEEEoverridecommandlockouts
\usepackage{cite}
\usepackage{hyperref}
\usepackage{multirow}
\usepackage{amsmath,amssymb,amsfonts}
\usepackage{graphicx}
\usepackage{textcomp}
\usepackage{xcolor}
\usepackage{comment}
\usepackage{graphicx}
\usepackage{array}
\usepackage{xcolor}
\usepackage{tabularx}

\usepackage{makecell}
\usepackage{mathtools}
\usepackage{rotating}

\usepackage{amsmath,amsfonts}
\usepackage{array}
\usepackage[caption=false,font=normalsize,labelfont=sf,textfont=sf]{subfig}
\usepackage{textcomp}
\usepackage{stfloats}
\usepackage{url}
\usepackage{verbatim}
\usepackage{graphicx}
\usepackage{algorithmicx}
\usepackage{algorithm}
\usepackage{algpseudocode}

\def\BibTeX{{\rm B\kern-.05em{\sc i\kern-.025em b}\kern-.08em
    T\kern-.1667em\lower.7ex\hbox{E}\kern-.125emX}}
\begin{document}

%%%%%%%%%%%%%%%%%%%%%%%%%%%%%%%%%%%%%%%%%%%%%%%%%%%%%%%%%%%%%%%%%%%%%%%%%%%%%%%%
%%%%%%%%%%%%%%%%%%%              TITLE SECTION              %%%%%%%%%%%%%%%%%%%%
%%%%%%%%%%%%%%%%%%%%%%%%%%%%%%%%%%%%%%%%%%%%%%%%%%%%%%%%%%%%%%%%%%%%%%%%%%%%%%%%
%\title{LC Resonant Clock Resource Minimization using Compensation and sizing Decoupling Capacitance}
\title{Integrated AHB to APB Bridge Using Raspberry Pi and Artix-7 FPGA}
%\title{Power and Skew Reduction Using Resonant Energy Recycling in 14-nm FinFET Clocks}

%\title{Low Power Cc-enhanced LC Resonant Clock Design Essentials}

%%%%%%%%%%%%%%%%%%%             AUTHORS SECTION             %%%%%%%%%%%%%%%%%%%%
\author{
%Anonymous for review purposes. Do not distribute.

Gopi Chand Ananthu,
 Riadul~Islam,~\IEEEmembership{Senior Member,~IEEE}
%<-this stops a space

\thanks{G. Ananthu and R. Islam are with the Department 
of Computer Science and Electrical Engineering, University of Maryland, Baltimore County, 
MD 21250, USA e-mail: {riaduli@umbc.edu}.}
%\thanks{B Saha is with the Si2Chip Technologies, 
%Road 1B, Gayatri Tech Park, Bengaluru, Karnataka 560066, India e-mail: {biprangshu.saha@si2chip.com}}
\thanks{}

\thanks{}
}
% The paper headers
\markboth{IEEE Transactions on Circuits and Systems--I}
%\markboth{Final semester project of Gopi Chand Ananthu}
%\markboth{IEEE Transactions on Computer-Aided Design of Integrated Circuits and Systems}%
{Shell \MakeLowercase{\textit{et al.}}: ??????}

\newcommand{\fixme}[1]{{\Large FIXME:} {\bf #1}}

% make the title area
\maketitle

%%%%%%%%%%%%%%%%%%%             ABSTRACT SECTION            %%%%%%%%%%%%%%%%%%%%
\begin{abstract}
This project focuses on the design and implementation of an AHB to APB Bridge for efficient communication in System-on-Chip (SoC) architectures. The Advanced High-performance Bus (AHB) is used for high-speed operations, typically connecting processors and memory, while the Advanced Peripheral Bus (APB) is optimized for low-power, low-speed peripheral devices. The AHB to APB Bridge serves as an interface that converts complex, high-speed AHB transactions into simpler, single-cycle APB transactions, enabling seamless data transfer between fast components and slower peripherals.
The bridge manages clock domain synchronization, transaction conversion, and flow control, ensuring compatibility between AHB’s burst transfers and APB’s non-pipelined protocol. Implemented in Verilog and simulated on FPGA using Xilinx Vivado, this bridge design provides a robust solution for integrating high-performance and low-power components within a single SoC. This project also evaluates the bridge’s functionality and performance through testbenches covering various operational scenarios, validating its efficiency in handling diverse system requirements. 

\end{abstract}
\section{Introduction}
\label{sec:intro}
The AHB to APB Bridge is a vital element in System-on-Chip (SoC) architectures, facilitating efficient communication between high-performance and low-power components. In modern SoCs, the Advanced High-performance Bus (AHB) is used for high-speed communication, connecting processors and memory, while the Advanced Peripheral Bus (APB) caters to lower-speed peripherals like timers, GPIOs, and UARTs. The AHB to APB Bridge ensures seamless integration by converting AHB's complex, high-speed burst transactions into APB's simpler, single-cycle transactions \cite{ARM_ahb:2010,ARM_apb:2010}.This bridge also handles synchronization across clock domains and manages transaction flow, which is critical for bridging fast AHB masters with slower APB slaves \cite{Wolf_soc:2015}. By supporting this transaction flow and ensuring correct handshaking mechanisms, the AHB to APB bridge becomes indispensable for designing scalable and power-efficient SoCs that balance performance and energy efficiency \cite{Patterson_computer:2014,Xilinx_vivado:2022}.

The AHB and APB are core components of ARM's Advanced Microcontroller Bus Architecture (AMBA) standard, which optimizes communication within SoC designs \cite{ARM_ahb:2010, ARM_apb:2010}. AHB supports high-speed, high-performance communication between CPUs, memory~\cite{islam2020resonant}, and other high-speed peripherals~\cite{ARM_ahb:2010}. APB, on the other hand, is designed for lower-speed, simpler peripherals, focusing on low-power data transfers \cite{ARM_apb:2010}. The AHB to APB Bridge enables these two buses to work in tandem by translating AHB's burst-mode transactions into APB's single-cycle, non-pipelined transactions \cite{Wolf_soc:2015}. This ensures efficient data exchange between these disparate components, balancing the system's performance and energy efficiency \cite{Xilinx_fpga:2020}.

The AHB to APB Bridge offers significant benefits in SoC designs. It enables seamless communication between high-speed AHB components and low-speed APB peripherals by converting burst-mode AHB transactions into simpler APB transactions \cite{ARM_ahb:2010, ARM_apb:2010}. The bridge ensures synchronization across different clock domains, preventing timing issues and ensuring reliable data transfer \cite{Saini_can:2024}. By facilitating integration with low-power APB peripherals, it supports efficient power management, making it ideal for energy-sensitive applications \cite{Xilinx_vivado:2022, Xilinx_fpga:2020}. Additionally, the bridge manages control signals for flow control and error handling, ensuring data integrity and preventing communication errors \cite{Harris_digital:2012, Patterson_computer:2014}. This capability allows for a mix of high-performance and low-power components to coexist, supporting scalable, flexible designs for a wide range of applications, from consumer electronics to automotive and industrial systems.

\subsection{Motivation}
\label{subsec:moti}
In modern SoC designs, the integration of high-performance processors and low-power peripheral devices is critical for achieving optimal performance and energy efficiency. Unlike conventional signaling techniques~\cite{islam2018hcdn, islam2011high, islam2016cmcs, islam2021high, islam2018dcmcs, guthaus2017current, islam2018negative, islam2018low}, AHB and APB operate on different protocols and speed requirements, which can create challenges for direct communication \cite{ARM_ahb:2010, ARM_apb:2010}. The AHB to APB Bridge provides a solution by enabling seamless communication between these components without compromising data integrity or overall system performance \cite{Wolf_soc:2015}.The development of this bridge facilitates flexible and scalable SoC designs, catering to applications ranging from mobile devices to industrial systems \cite{Patterson_computer:2014, Xilinx_fpga:2020}. The motivation behind this project is to bridge the communication gap between high-speed and low-power components, thereby advancing efficient and versatile SoC designs \cite{rahi2024system, islam2022early, guo202468, qiu2024vsoc, Saini_can:2024, smith2024mkidgen3, islam2019low, islam2022feasibility}. Building this design on the reconfigurable design can help the researchers to use this design in there research to communicate between the components.

\subsection{Contributions}
\label{subsec:org}
This project employs an efficient design to transmit input signals from a Raspberry Pi to an Artix-7 FPGA using an SPI interface. The Raspberry Pi sequentially sends serialized input signals bit by bit over SPI, which are then decoded into parallel inputs by the FPGA \cite{Hossain_spi:2015}. A Finite State Machine (FSM) generates control signals for multiplexers and demultiplexers on the FPGA to ensure proper routing and synchronization of data \cite{Wolf_soc:2015}. The multiplexer directs SPI Slave outputs to the FPGA's processing modules, while the demultiplexer gathers inputs for hardware logic \cite{Harris_digital:2012}. The processed outputs from the FPGA are sent back through the SPI interface to the Raspberry Pi for validation and visualization. Python scripts on the Raspberry Pi manage SPI communication, data encoding, and output analysis, ensuring seamless integration \cite{Jain_digital:2010}. This methodology optimizes pin usage on the FPGA through serialization and provides robust synchronization, enabling efficient processing and communication \cite{Saini_can:2024}.

%existing in-memory architecture example and still could not resolve power issue. We resolve this by introducing resonant

% what are different aspect of our in-memory and major contributions

\section{Background}
\label{sec:background}

The AHB to APB Bridge is a component in SoC architectures that enables communication between the high-speed  \cite{Ali_rpi_fpga:2021}AHB and the low-speed APB . It acts as an interface, converting complex AHB transactions into simpler APB transactions, allowing high-performance components like CPUs and memory to interact seamlessly with low-power peripheral devices, such as UARTs and GPIOs, within a single SoC 

The AHB  to APB  Bridge is a fundamental component in modern SoC architectures, enabling seamless communication between high-performance and low-power subsystems \cite{ARM_ahb:2010, ARM_apb:2010}. The AMBA standard, developed by ARM, defines both AHB and APB protocols to address the growing demands for high-speed processing and efficient peripheral management. The AHB, as a high-bandwidth bus, supports burst-mode transfers, pipelined operations, and multi-cycle data processing, making it ideal for connecting high-speed components such as processors, memory controllers, and DMA engines \cite{Patterson_computer:2014}. Conversely, the APB is designed for simplicity, offering single-cycle, low-power transactions to efficiently interface with peripherals such as GPIOs, timers, UARTs, and sensors \cite{Jain_digital:2010}.

The AHB to APB Bridge is widely deployed in diverse applications requiring efficient communication between high-speed processors and low-speed peripherals in SoCs. In embedded systems, it supports communication between processing cores and peripherals like sensors and timers, making it essential for microcontrollers and development boards . In consumer electronics, the bridge connects fast processors to audio controllers, display drivers, and I/O interfaces, enabling efficient operation in devices like smartphones and smart home systems . In automotive systems, it facilitates interaction between CPUs and control units, powering applications like infotainment and engine management \cite{Xilinx_fpga:2020, Tessier_reconfigurable:1998}. In industrial automation, it links high-speed controllers with sensors and actuators, ensuring precise control in manufacturing and robotics \cite{Xilinx_vivado:2022}. Additionally, in networking, the bridge connects processors with Ethernet controllers and management interfaces, ensuring data flow in routers and switches \cite{Trimberger_fpga:2015}. Overall, the bridge is fundamental to integrating high-performance processors with diverse peripherals, supporting efficient, scalable, and power-optimized designs.

The AHB to APB Bridge plays a critical role in protocol translation, converting AHB’s complex burst-mode and pipelined operations into APB’s non-pipelined, single-cycle transactions \cite{Wolf_soc:2015}. This bridge also resolves clock domain mismatches since the AHB operates at higher frequencies for performance-critical tasks, while the APB runs at lower frequencies to save power \cite{Xilinx_vivado:2022}. By handling address decoding, control signal generation, and transaction synchronization, the bridge enables the smooth integration of low-speed peripherals with high-speed AHB components \cite{Trimberger_fpga:2015}. This integration ensures power efficiency, scalability, and modularity in SoC designs, making the AHB to APB Bridge a critical enabler for applications ranging from embedded systems and mobile devices to industrial automation and automotive electronics

%{\color{blue}

\begin{comment}
    The resonant write driver, shown in Figure~\ref{fig:gsr_tank}~(b), utilizes an inductor ``$L$" to store this discharged energy and recycles it back during the next precharge phase~\cite{Islam_sram:2021, dhandeep_23_rcim}. During resonance operation, the ``$vsr$" signal initially goes high, allowing the inductor to store the dissipated energy from the bitline. Subsequently, the ``$vdn$" signal is asserted to fully discharge the bitline. In the next precharge phase, the ``$vsr$" signal is enabled again to recycle back the energy stored in the inductor into the bitline.
\end{comment}

%}

\section{Proposed Methodology}

%%% introduction to the proposed methodology

The AHB to APB Bridge using the Raspberry Pi and Artix-7 FPGA is designed to facilitate seamless communication between high-speed AHB components and low-power APB peripherals. The project integrates data transfer, processing, and protocol conversion using various functional modules. The system uses SPI communication for data transmission between the Raspberry Pi and FPGA. The following subsections describe the architecture and the signals involved in this design \cite{Saini_can:2024}.

\subsection{System Architecture and Module Breakdown}

\begin{figure}[t!]
\begin{center}
\includegraphics[width = 0.5\textwidth]{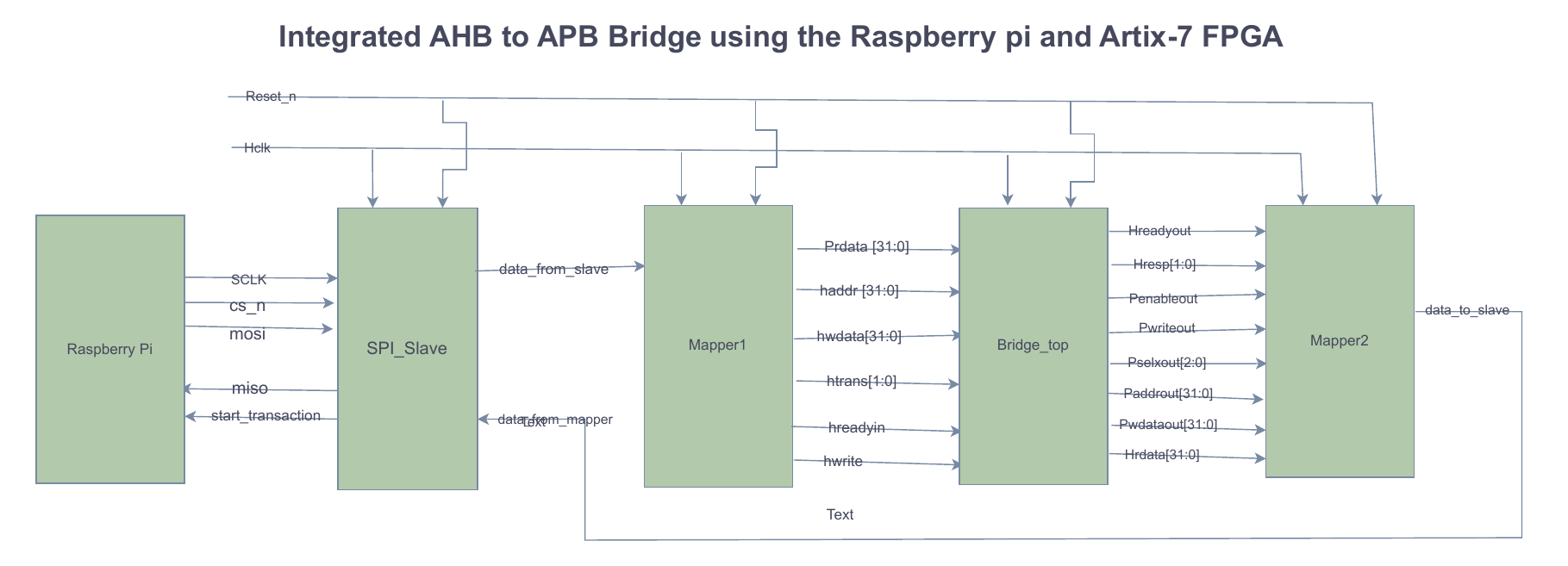}
\end{center}
\vspace{-0.50cm}
\caption{Integrated AHB to APB Bridge using the Raspberry pi and Artix-7.}
\label{fig:10t_cell_layout}
%\vspace{-0.50cm}
\end{figure}

%{\color{blue}
Figure~\ref{fig:10t_cell_layout}(a) The figure  illustrates the architecture of the AHB to APB Bridge, showcasing the flow of signals between the Raspberry Pi, SPI\_Slave module, Mapper modules, and the Bridge\_top. The design highlights the systematic protocol conversion process for seamless communication between high-speed and low-power components.
%}

The project architecture consists of five key modules: Raspberry Pi(Master Device), SPI Slave(Data Reception Module), Mapper1(Signal Mapping Module), BridgeTop(AHB to APB Conversion Module).

\subsubsection{Raspberry Pi}
\label{subsec:features}
The Raspberry Pi acts as the master device, generating the 100-bit wide input data.The input signals are serialized and sent to the FPGA via the MOSI line using SPI communication.The Raspberry Pi also receives processed output data bit by bit through the MISO line for validation and visualization.

\subsubsection{SPI slave}
\label{subsec:rf}
This module receives serialized data (100 bits) from the Raspberry Pi over the SPI\cite{Das_spi_fpga:2021} interface. \cite{Reddy_fpga_ahb:2023}It handles data reception, synchronization and outputs a start transaction signal to trigger further processing.\cite{Nayak_fpga:2019}\cite{Das_spi_fpga:2021}\cite{Chen_rpi_fpga:2023} It sends the received 1 bit data sequentially to Mapper1 and outputs control signals such as start transaction to indicate valid data reception \cite{Hossain_spi:2015}\cite{Sharma_artix7:2022}.

\subsubsection{Mapper1}
\label{subsec:features}
The Mapper1 module takes the 1-bit input from the SPI Slave and assembles it into parallel 100-bit data.\cite{Saxena_ahb_apb:2022} It maps sections of this 100-bit data to specific AHB-compatible signals: prdata, haddr, hwdata, htrans, hreadyin, and hwrite. These outputs are sent to the Bridge top module for further processing \cite{Saini_can:2024}.

\subsubsection{Bridge Top}
\label{subsec:features}
This module forms the core of the project and handles protocol conversion between AHB and APB. It processes the AHB-compatible signals from Mapper1, converts them into APB-compatible control signals, and manages read/write operations. Outputs include APB signals like Pwriteout, Penableout, Pselxout, Pwdataout, and Paddrout \cite{Saini_can:2024}.

\subsubsection{Mapper2}
\label{subsec:features}
The Mapper2 module gathers outputs from the Bridge top module, including APB signals, response flags, and status bits. These signals are aggregated into a 104-bit data output and sent back to the Raspberry Pi, one bit at a time, using the MISO line. This output ensures that the Raspberry Pi receives all processed information for verification 

\subsection{AHB to APB Bridge Architecture}
\label{subsec:features}
The Bridge Top Architecture serves as the core module for converting AHB signals into APB-compatible signals.\cite{Gandhi_amba:2020}\cite{Prasad_bridge:2021} It consists of three key interconnected modules: AHB Slave Interface, APB FSM Controller, and APB Interface. \cite{Mehta_area_timing:2023}The architecture ensures protocol conversion,\cite{Joshi_icc2:2021} control signal management, and proper synchronization between high-speed AHB and low-power APB domains \cite{ARM_ahb:2010,ARM_apb:2010}.

The Bridge Top Architecture consists of three key modules. They are AHB slave Interface , APB FSM Controller,APB Interface

\begin{figure}[h]
\begin{center}
\includegraphics[width = 0.5\textwidth]{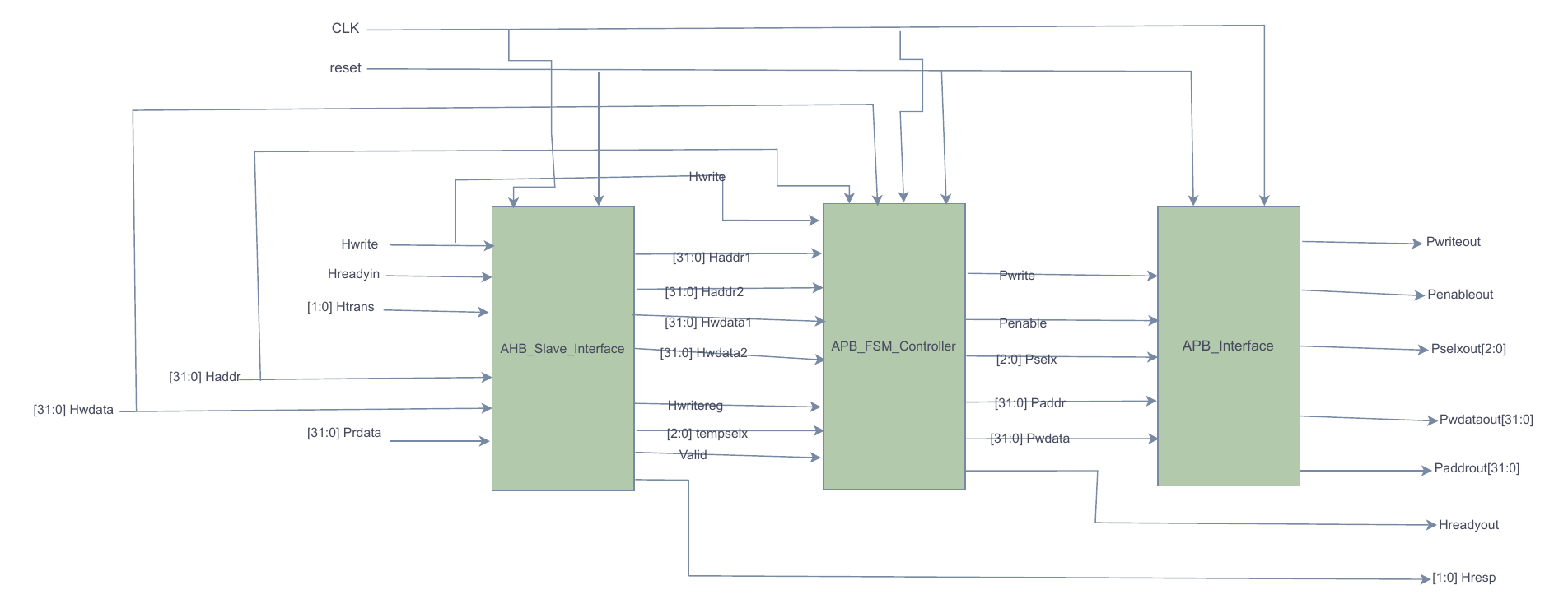}
\end{center}
\caption{AHB to APB Bridge.}
\label{fig:novel_circuit}

\end{figure}

%%% overview of performing NAND/NOR and generation of rwl pulse width

\subsubsection{AHB Slave Interface}
The AHB Slave Interface acts as the AHB slave module. It receives AHB signals such as Haddr, Hwrite, Hreadyin, Htrans, and Hwdata from the AHB master. It processes these inputs, validates the transaction, and generates control signals like Hwritereg, Valid, and tempselx to be used by the subsequent module. \cite{Kumar_ahb2apb:2018}It also pipelines AHB addresses and data to ensure smooth handling of read and write operations \cite{ARM_ahb:2010}.

\subsubsection{APB FSM Controller}
The APB FSM Controller is responsible for managing APB transactions. It receives control signals like Hwritereg and Valid from the AHB Slave Interface and generates APB control signals such as Pwrite, Penable, and Pselx. Using a FSM, it orchestrates the flow of read and write operations, ensuring proper handshaking between the AHB and APB domains \cite{ARM_apb:2010}.

\subsubsection{APB Interface}
The APB Interface generates APB-compatible signals required to communicate with APB peripherals. It outputs Pwriteout, Penableout, Pselxout, Pwdataout, and Paddrout, which are sent to the selected APB peripheral. Additionally, it manages transaction completion using Hreadyout and generates response signals like Hresp to communicate back with the AHB master \cite{Jain_digital:2010}.

\subsection{Signal Description and Operations}

Each module in the system communicates using well-defined signals to ensure synchronized and robust operation. Below is a detailed explanation of the signals and their roles:\\

\subsubsection{SPI slave signals}
Inputs for the SPI Slave are clk, resetn, mosi, sclk, and csn. Outputs from the SPI Slave include miso, starttransaction, and data to Mapper1. The SPI Slave uses the SPI clock (sclk) to sample input data (mosi) bit by bit.\cite{Khan_ahb_fpga:2022} Once the 100 bits are received, it triggers the starttransaction signal and sends the data sequentially to Mapper1 \cite{Hossain_spi:2015}.

\subsubsection{Mapper1 Signals}
Inputs for Mapper1 are clk, resetn, and data from Slave. Outputs from Mapper1 include Prdata, Haddr, Hwdata, Htrans, Hreadyin, and Hwrite. The Mapper1 module accumulates 100-bit data from the SPI Slave and maps sections of this data into AHB-compatible signals. These signals are sent to the Bridge Top module to initiate AHB transactions \cite{Wolf_soc:2015}.

\subsubsection{BridgeTopSignals}
Inputs of BridgeTopSignals are Haddr, Hwdata, Htrans, Hreadyin, Hwrite.And the outputs of the Bridge is Pwriteout, Penableout, Pselxout, Pwdataout, Paddrout.The Bridge top module converts AHB protocol signals into APB protocol signals. It generates the necessary control signals (Pwriteout, Penableout) and routes the address/data to the appropriate APB peripheral.\cite{ARM_ahb:2010}.

\subsubsection{Mapper2Signals}
Inputs of Mapper2 are Pwriteout, Penableout, Pselxout, Pwdataout, Paddrout. Outputs of the module is data\_to\_slave.Mapper2 combines APB outputs into a 104-bit data structure. This data is serialized and sent bit by bit to the Raspberry Pi via the MISO line.\cite{Harris_digital:2012}

\subsubsection{AHB\_Slave\_Interface Signals}
Inputs of the AHB Slave Interface are clk, reset, Haddr, Hwdata, Htrans, Hwrite, Hreadyin, and Prdata. Outputs of the module include Haddr1, Haddr2, Hwdata1, Hwdata2, Hwritereg, and tempselx \cite{ARM_ahb:2010}.

\begin{figure}[h]
\begin{center}
\includegraphics[width = 0.5\textwidth]{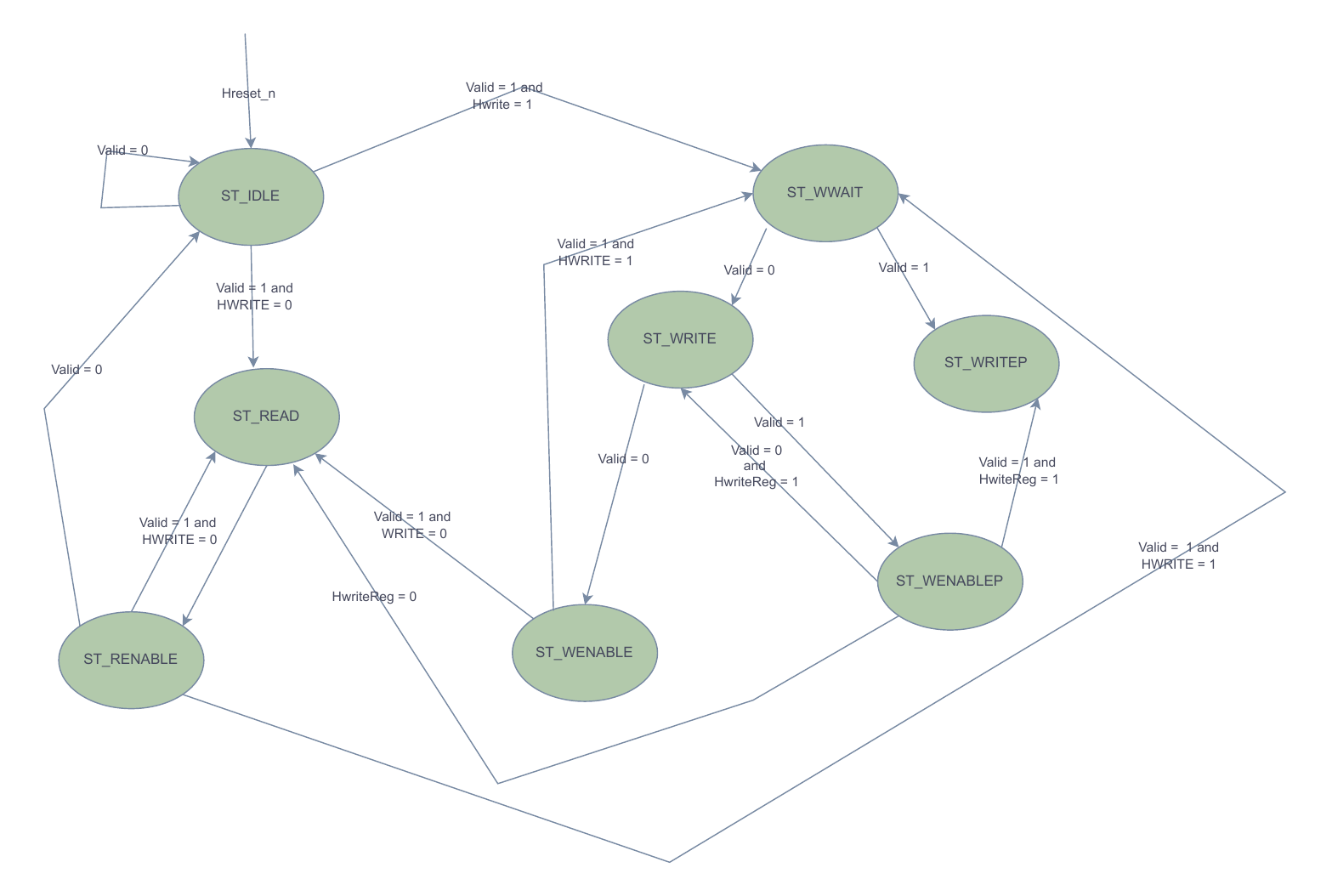}
\end{center}
\caption{Finite State Machine.}
\label{fig:novel_circuit}

\end{figure}

\subsubsection{APB\_FSM\_Controller Signals}
Inputs of the APB FSM Controller are Hwritereg, Valid, Haddr1, Hwdata1, Haddr2, Hwdata2, and tempselx. Outputs of the module include Pwrite, Penable, Pselx, Paddr, and Pwdata \cite{ARM_apb:2010}.

\subsubsection{APB\_Interface Signals}
Inputs of the APB Interface are Pwrite, Penable, Pselx, Paddr, and Pwdata. Outputs of the module include Pwriteout, Penableout, Pselxout, Pwdataout, Paddrout, Hreadyout, and Hresp.\cite{ARM_apb:2010}

%%% working of sense amplifier circuit with reference to transient simulations

%\caption{Functionality waveform illustrating NAND computation with "01"/"10" data and a successful energy-recycling writeback operation of the result.}

%\caption{The SPICE simulation confirms the correct in-memory computation considering logical NAND2 operations with ``01/10" data and a conventional energy-recycling writeback operation.}
%\label{fig:functionality}

%%% resonant write driver overview, and how the flow after computation is done

%%% detailed description of resonant write driver

%}

\section{Experiments and Results}
\label{sec:exp}
\subsection{Experimental Setup}

The experimental setup for the implementation of the AHB to APB bridge consists of the Artix-7 100TCSG324 FPGA \cite{Trimberger_fpga:2015} and Raspberry Pi 4 Model B \cite{raspberrypi4_datasheet}. Similar to conventional approaches~\cite{saini2024reconfigurable, Croteau:2024, Kiriakidis:2022}, the FPGA serves as the processing platform for the implementation of the bridge, hosting Verilog modules for SPI communication, data mapping, protocol conversion, and signal aggregation. The Raspberry Pi acts as the master device, transmitting serialized 100-bit input data to the FPGA \cite{Sharma_artix7:2022} via the SPI interface and receiving the processed 104-bit output bit by bit. The design is developed, simulated, and implemented using the Xilinx Vivado Design Suite \cite{Xilinx_vivado:2022}, which facilitates RTL coding, behavioral simulation, and hardware synthesis. Additionally, Python scripts running on the Raspberry Pi manage SPI communication, including data serialization, transmission, and output validation.To optimize and evaluate the design for area, power, and timing, the Synopsys Design Compiler (DC) is utilized for synthesis, providing detailed reports on gate-level netlist area, power consumption, and timing delays. Post-synthesis analysis ensures that the bridge meets the required timing constraints and achieves efficient resource utilization. The Synopsys IC Compiler II (ICC2) is used for physical implementation, enabling place-and-route operations while refining the design for reduced power and optimized area. Together, the Vivado, Design Compiler, and ICC2 tool chains provide a comprehensive framework for RTL-to-GDSII implementation, ensuring synchronized communication, efficient hardware processing, and accurate verification of bridge functionality.

\subsection{Experimental Results}
The waveform for the Bridge Top module illustrates the communication between the AHB and APB interfaces\cite{ARM_ahb:2010,ARM_apb:2010}. Initially, the system is in an idle state, with clk running continuously and reset de-asserted. The AHB interface begins a transaction when Htrans transitions to Non-sequential (value 3), indicating a valid transfer. During this phase, the AHB address (Haddr) is updated to 0x8000000C and 0x80000008, and write data (Hwdata) holds values like 0xFFFFFFFF.

\begin{figure}[b!]
\centerline{\includegraphics[width = 0.5\textwidth]{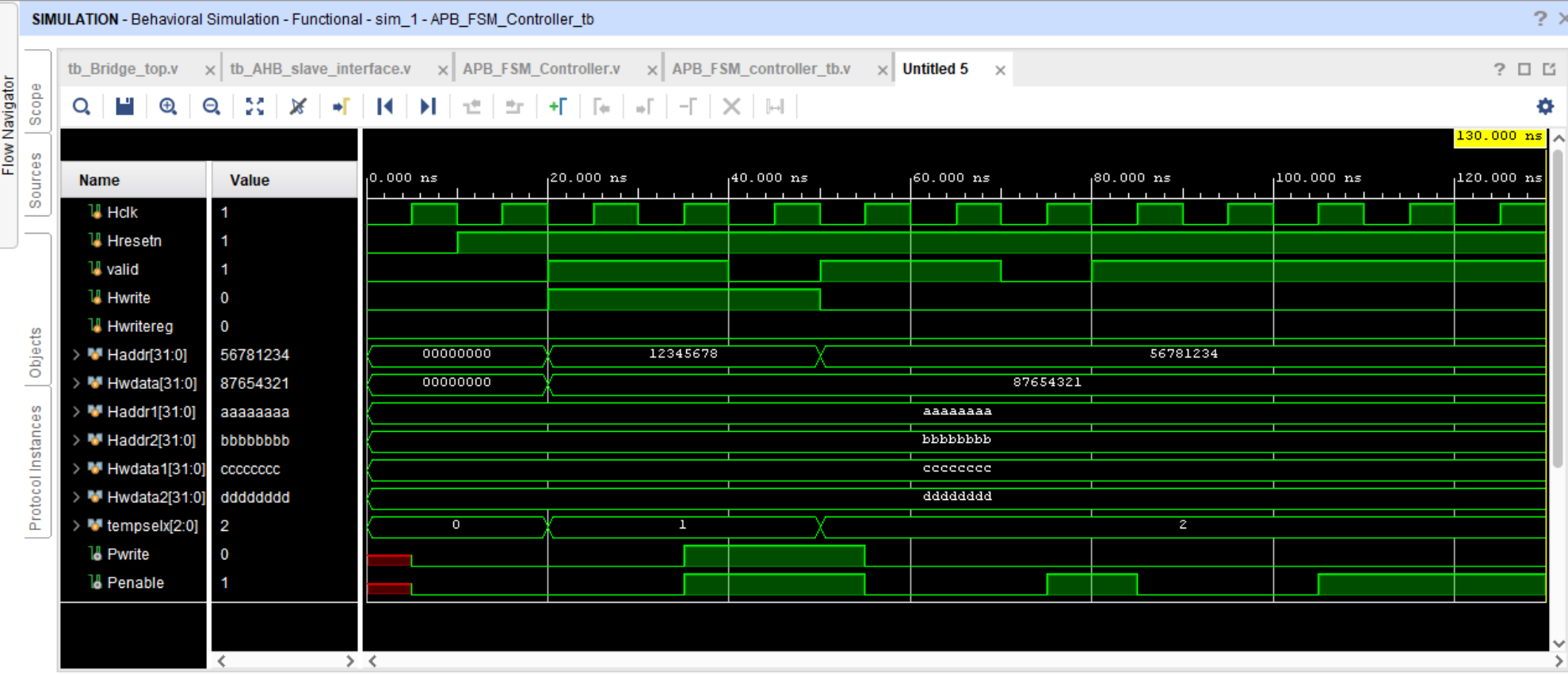}}
\caption {Simulation result for APB FSM Controller shows the output signals Pwrite,Penable  high after the Hwrite and Valid goes high.}
\label{fig:ahb_fsm_page1}
%\vspace{-0.3cm}
\end{figure}

Fig.4 presents the simulation results of the Finite State Machine (FSM) for the BridgeTop module, demonstrating signal transitions during its functional operation. The clock signal (Hclk) serves as the timing reference, generating a stable periodic waveform that drives synchronous events. At the start of the simulation, the reset signal (Hresetn) is asserted low and later de-asserted high, releasing the FSM from reset and initiating its normal operation. The valid signal asserts high, indicating that valid data is being transmitted.
The input address bus (Haddr[31:0]) transitions to 0x12345678 and 0x56781234, denoting address values sent during specific clock cycles, while the write data bus (Hwdata[31:0]) carries the value 0x87654321 during write operations. The Hwrite signal, initially low, toggles high to indicate an active write operation. Similarly, the tempselx[2:0] signal transitions from 0 to 1 and later to 2, showcasing the selection of different states or modules during the FSM operation. The Pwrite and Penable signals are asserted high, signaling the enablement of peripheral write operations in the APB interface.
The simulation results validate the proper operation of the FSM, including synchronized state transitions, address decoding, and data transfer between the AHB and APB\cite{Kumar_ahb2apb:2018} domains. The waveforms confirm accurate signal generation and control, ensuring seamless communication between the system components.

\begin{figure}[t!]
\centerline{\includegraphics[width = 0.5\textwidth]{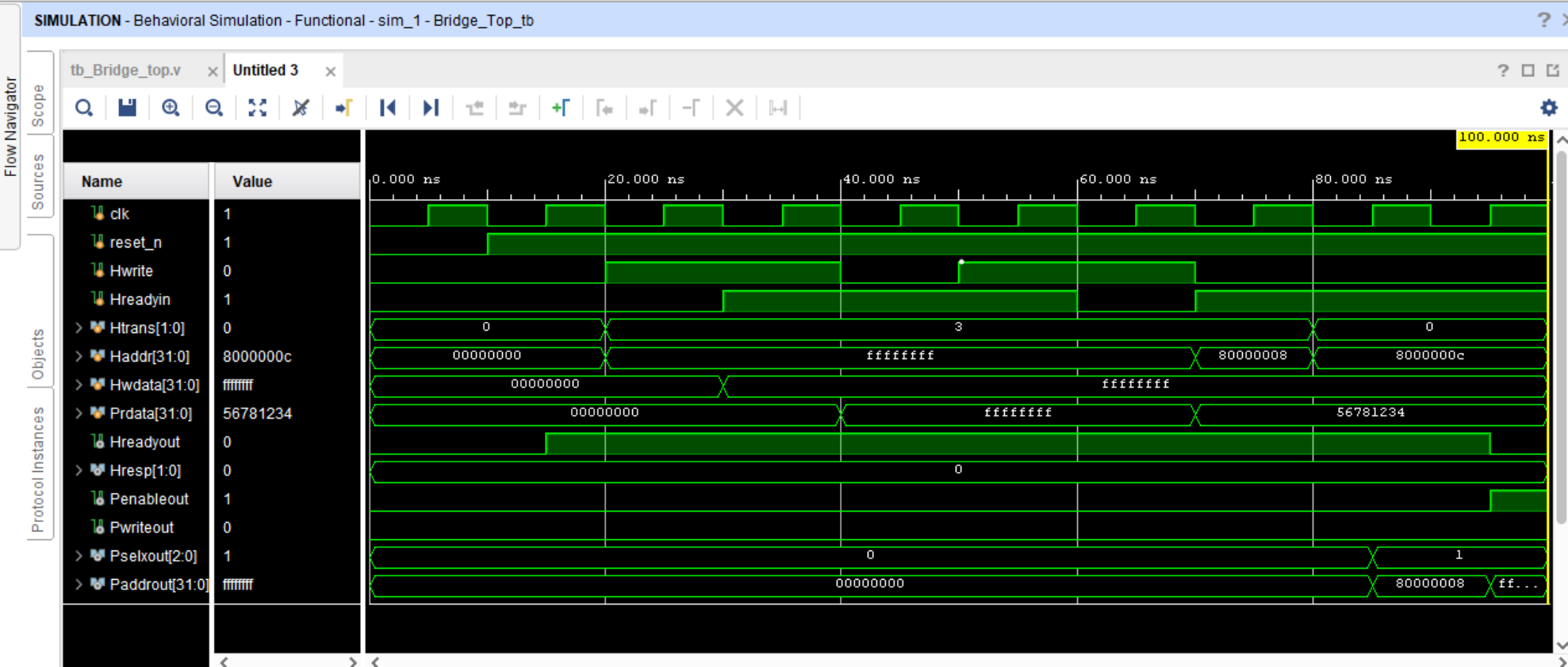}}
\caption {Simulation result for AHB to APB.}
\label{fig:ahb_fsm_page1}
%\vspace{-0.3cm}
\end{figure}

Fig.5 presents the simulation results for the AHB to APB Bridge, demonstrating the correct operation of the bridge during read and write transactions. The clock signal (clk) remains stable with a periodic waveform, ensuring synchronization across the design. The reset signal (resetn) is asserted initially and de-asserted after stabilization, enabling normal operation of the system. The Hwrite signal is set to 1, indicating a write operation, while Hreadyin remains asserted, signaling that the AHB master is ready to initiate transactions. The Htrans[1:0] value changes to 3 during the valid phase, representing a non-sequential transfer.
The Haddr[31:0] signal outputs the address 0x8000000C, which is captured and routed through the AHB interface. Concurrently, Hwdata[31:0] carries the data 0xFFFFFFFF, representing the write data being transmitted to the APB slave. The APB side reflects this transaction with the Paddrout[31:0] showing the address 0x8000000C, confirming correct address decoding. Additionally, the Pwriteout signal asserts 1, validating the ongoing write transaction, and Penableout is asserted, completing the APB transfer phase.
During the read cycle, Prdata[31:0] carries the data 0x56781234, signifying successful data retrieval from the APB slave. The Hresp[1:0] signal remains 0, indicating no errors occurred during the communication. The Pselxout[2:0]\cite{Kumar_ahb2apb:2018} reflects appropriate slave selection values as 1, ensuring correct routing to the intended APB peripheral.
This simulation waveform confirms the bridge’s proper operation, including accurate protocol conversion, address mapping, and data transfer between the AHB master and APB slave. Both read and write transactions are successfully synchronized, validating the seamless communication enabled by the bridge.

\

\begin{figure}[h]
\begin{center}
\includegraphics[width = 0.5\textwidth]{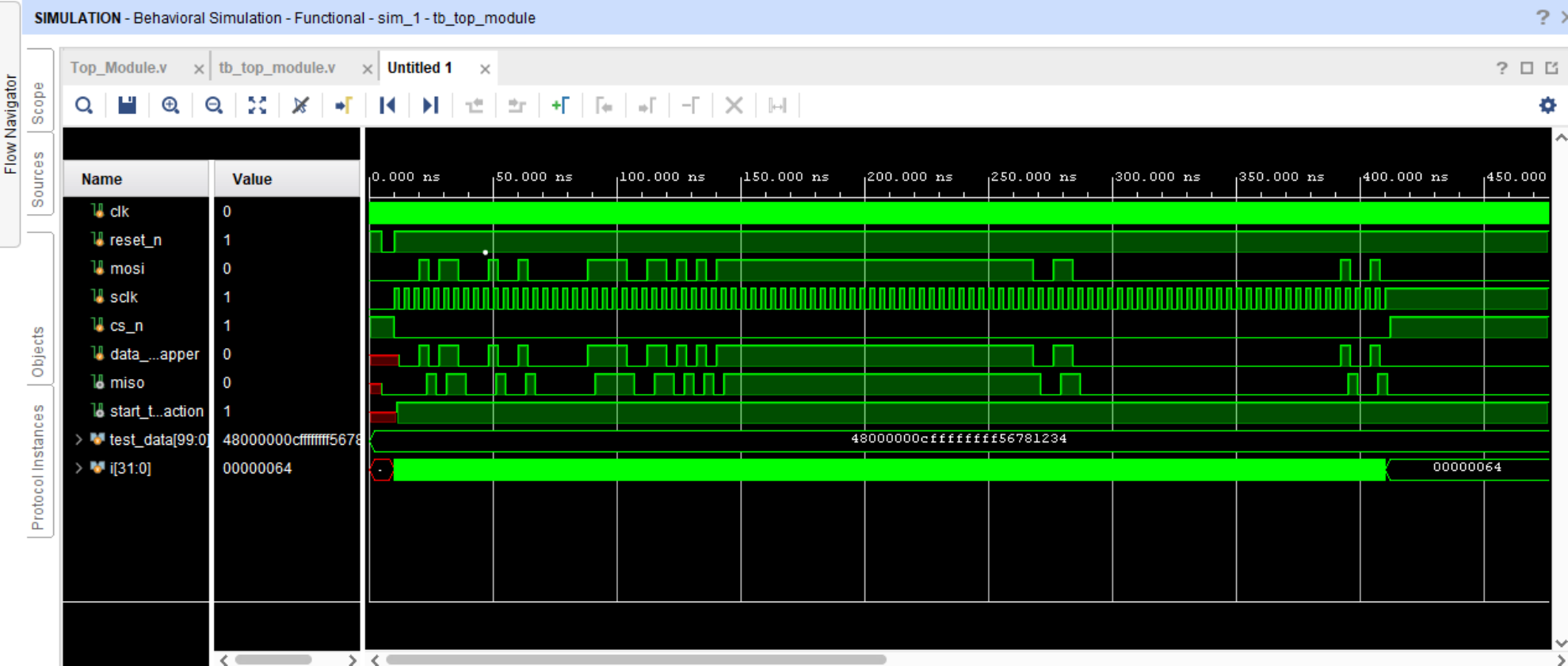}
\caption {Simulation result for Integrated  AHB to APB Bridge}
\end{center}
\end{figure}

Figure 6 illustrates the simulation results for the AHB to APB Bridge, showcasing signal transitions and the interaction between AHB and APB buses. The clock (clk) remains active throughout the simulation, ensuring synchronous operation. The reset signal (resetn) is asserted initially to initialize the system and is subsequently deasserted to begin normal operation. The Hwrite signal indicates a write operation, while the Hreadyin signal is asserted, enabling the transfer process.
The AHB transaction begins with the Haddr signal set to 8000000C and Hwdata carrying the value FFFFFFFF. Concurrently, Htrans[1:0] is toggled to 3, signaling a non-sequential transfer. The Paddrout[31:0] and Pselxout[2:0] signals reflect the APB address and selection outputs, respectively, confirming the translation of the AHB signals into the APB domain. Penableout is asserted, enabling the APB transfer, while Pwriteout remains 1, indicating a write operation.
The Prdata[31:0] signal outputs the value 56781234 at a later stage, signifying the readback or response data from the peripheral. The data transfer completes successfully with synchronization between the AHB and APB signals, as indicated by the stable Hresp[1:0] signal.
Figure 7 presents the simulation waveform of the top module interfacing an SPI master and FPGA hardware. The clk remains low initially, transitioning to an active state as the simulation progresses. The mosi (Master Out Slave In) and miso (Master In Slave Out) signals demonstrate the serialized SPI communication, where the sclk (SPI clock) ensures proper synchronization. The csn (chip select) signal toggles appropriately to enable data transfer.
The test data input (testdata[99:0]) is transmitted in multiple chunks, starting with the 100-bit value 48000000CFFFFFFFFF56781234. The corresponding 32-bit index (I[31:0]) increments and outputs 00000064, validating the received data. The successful handshake between the SPI master and the FPGA\cite{Chen_rpi_fpga:2023, Das_spi_fpga:2021} is evident from the signal transitions, with the $start_t$ action signal indicating the initiation of data capture and processing.
These results confirm the correct functionality of the AHB to APB Bridge and SPI interface, demonstrating reliable data transfer, signal routing, and synchronization across different domains. This verification ensures the design's robustness for practical implementation.

\begin{table}[h!]
\centering
\caption{Raspberry PI Input and FPGA Output}
\label{tab:raspberry_pi_results}
\resizebox{\linewidth}{!}{%
\begin{tabular}{|l|l|}
\hline
\textbf{Input to FPGA (From Raspberry Pi)} & \textbf{Output From FPGA}      \\ \hline
Prdata (Hex, 31:0): 0x12345678             & Hrdata (31:0): 0x12345678      \\ \hline
Haddr (Hex, e.g., 0x8C000000): 0x8C000000 & Paddr (31:0): 0x8C000000       \\ \hline
Hwdata (Hex, 0x87654321): 0x87654321       & Pwdata (31:0): 0x87654321      \\ \hline
Htrans[1:0] (Binary, e.g., 10): 10         & Pselx (2:0): 0101              \\ \hline
Hreadyin[0] (Binary, e.g., 1): 1           & Hresp (1:0): 0b10              \\ \hline
Hwrite[1] (Binary, e.g., 1): 1             & Hreadyout: 1                   \\ \hline
-                                          & Pwrite: 1                      \\ \hline
-                                          & Penable: 1                     \\ \hline
\end{tabular}%
}
\end{table}

\begin{comment}
\begin{table*}[t]
\centering
\caption{Synthesis Area Report for \texttt{Bridge\_Top} design.}
\label{tab:area_report}
\resizebox{0.3\textwidth}{!}{%
\begin{tabular}
\hline
\textbf{Parameter}                              & \textbf{Value}         \\ 
\hline
Number of ports                                 & 206                    \\ 
Number of nets                                  & 453                    \\ 
Number of cells                                 & 352                    \\ 
Number of combinational cells                   & 114                    \\ 
Number of sequential cells                      & 238                    \\ 
Number of macros/black boxes                    & 0                      \\ 
Number of buf/inv                               & 26                     \\ 
Number of references                            & 19                     \\ 
\hline
Combinational area (units)                      & 54.612001              \\ 
Buf/Inv area (units)                            & 6.482400               \\ 
Noncombinational area (units)                   & 253.612809             \\ 
Macro/Black Box area (units)                    & 0.000000               \\ 
Net Interconnect area (units)                   & 477.019164             \\ 
\hline
\textbf{Total cell area (units)}                & \textbf{308.224810}    \\ 
\textbf{Total area (units)}                     & \textbf{785.243974}    \\ 
\hline
\end{tabular}}

\end{table*}
\end{comment}

\begin{table*}[t]
\centering
\caption{Synthesis Area Report for \texttt{Bridge\_Top} design.}
\label{tab:area_report}
\resizebox{0.3\textwidth}{!}{%
\begin{tabular}{|l|r|} % Added column specification
\hline
\textbf{Parameter}                              & \textbf{Value}         \\ 
\hline
Number of ports                                 & 206                    \\ 
Number of nets                                  & 453                    \\ 
Number of cells                                 & 352                    \\ 
Number of combinational cells                   & 114                    \\ 
Number of sequential cells                      & 238                    \\ 
Number of macros/black boxes                    & 0                      \\ 
Number of buf/inv                               & 26                     \\ 
Number of references                            & 19                     \\ 
\hline
Combinational area (units)                      & 54.612001              \\ 
Buf/Inv area (units)                            & 6.482400               \\ 
Noncombinational area (units)                   & 253.612809             \\ 
Macro/Black Box area (units)                    & 0.000000               \\ 
Net Interconnect area (units)                   & 477.019164             \\ 
\hline
\textbf{Total cell area (units)}                & \textbf{308.224810}    \\ 
\textbf{Total area (units)}                     & \textbf{785.243974}    \\ 
\hline
\end{tabular}%
}
\end{table*}

\begin{table*}[t]
\centering
\caption{Power Analysis Report for \texttt{Bridge\_Top} design.}
\label{tab:power_analysis_bridge}
\resizebox{1.0\textwidth}{!}{%
\begin{tabular}{|c|c|c|c|c|c|}
\hline
\textbf{Power Group} & \textbf{Internal Power (uW)} & \textbf{Switching Power (uW)} & \textbf{Leakage Power (pW)} & \textbf{Total Power (uW)} & \textbf{\% Contribution} \\ 
\hline
\textbf{io\_pad}          & 0.0000             & 0.0000             & 0.0000               & 0.0000             & 0.00\%  \\ 
\textbf{memory}           & 0.0000             & 0.0000             & 0.0000               & 0.0000             & 0.00\%  \\ 
\textbf{black\_box}       & 0.0000             & 0.0000             & 0.0000               & 0.0000             & 0.00\%  \\ 
\textbf{clock\_network}   & 0.0000             & 0.0000             & 0.0000               & 0.0000             & 0.00\%  \\ 
\textbf{register}         & 334.8159           & 1.8795             & 7.1027e+04           & 336.7664           & 97.70\% \\ 
\textbf{sequential}       & 0.0000             & 0.0000             & 0.0000               & 0.0000             & 0.00\%  \\ 
\textbf{combinational}    & 0.8345             & 7.0921             & 1.4948e+04           & 7.9415             & 2.30\%  \\ 
\hline
\textbf{Total}            & \textbf{335.6504}  & \textbf{8.9716}    & \textbf{8.5975e+04}  & \textbf{344.7079}  & \textbf{100.00\%} \\ 
\hline
\end{tabular}%
}
\end{table*}

\begin{table*}[t]
\centering
\caption{Timing Report for \texttt{Bridge\_Top} Design.}
\label{tab:timing_report}
\resizebox{1.0\textwidth}{!}{%
\begin{tabular}{|l|l|l|l|}
\hline
\textbf{Parameter}                       & \textbf{Description}                   & \textbf{Incremental Delay (ns)} & \textbf{Total Delay (ns)} \\ \hline
\textbf{Startpoint}                      & \texttt{reset\_n} (input port)         & -                               & -                         \\ \hline
\textbf{Endpoint}                        & \texttt{FSM\_PRESENT\_STATE\_reg\_1\_} & -                               & -                         \\ \hline
\textbf{Clock Source}                    & \texttt{clk}                           & 0.00                            & 0.00                      \\ \hline
\textbf{Clock Network Delay (ideal)}     & Ideal                                  & 0.00                            & 0.00                      \\ \hline
\textbf{Input External Delay}            & Input Delay                            & 0.10                            & 0.10                      \\ \hline
\texttt{U97/X}                           & Buffer (\texttt{SAEDRVT14\_BUF\_S\_0P75}) & 0.04                            & 0.14                      \\ \hline
\texttt{U127/X}                          & AND Gate (\texttt{SAEDRVT14\_AN3\_0P5})  & 0.03                            & 0.17                      \\ \hline
\texttt{U130/X}                          & AND Gate (\texttt{SAEDRVT14\_AN3\_0P5})  & 0.02                            & 0.19                      \\ \hline
\texttt{U102/X}                          & AND Gate (\texttt{SAEDRVT14\_AN3\_0P5})  & 0.04                            & 0.23                      \\ \hline
\texttt{U202/X}                          & Inverter (\texttt{SAEDRVT14\_INV\_0P5})  & 0.03                            & 0.26                      \\ \hline
\texttt{U206/X}                          & Complex Gate (\texttt{SAEDRVT14\_OAI22\_0P5}) & 0.02                            & 0.28                      \\ \hline
\texttt{U207/X}                          & Complex Gate (\texttt{SAEDRVT14\_AO221\_0P5}) & 0.02                            & 0.30                      \\ \hline
\textbf{Data Arrival Time}               & From Start to Endpoint                 & -                               & \textbf{0.30}             \\ \hline
\textbf{Clock Period}                    & Clock Cycle                            & -                               & 0.72                      \\ \hline
\textbf{Clock Uncertainty}               & Delay Variation                        & -0.07                           & 0.65                      \\ \hline
\textbf{Library Setup Time}              & Setup Margin                           & -0.01                           & 0.64                      \\ \hline
\textbf{Data Required Time}              & Total Required Time                    & -                               & \textbf{0.64}             \\ \hline
\textbf{Slack (MET)}                     & Timing Margin                          & -                               & \textbf{0.34}             \\ \hline
\end{tabular}%
}
\end{table*}

 Table ~\ref{tab:raspberry_pi_results} illustrates the interaction between the Raspberry Pi and FPGA by detailing the input signals sent to the FPGA and the corresponding output signals generated.
On the input side, the Raspberry Pi sends the Prdata value of 0x12345678, representing the read data, and an Haddr signal of 0x8C000000, which specifies the AHB bus address. Additionally, Hwdata (0x87654321) is provided as write data. The Htrans[1:0] signal is set to 10, indicating a non-sequential transfer, while Hreadyin[0] is asserted as 1, confirming the slave's readiness. The Hwrite[1] signal, also set to 1, specifies that a write operation is occurring.\cite{Narasimha_fpga_rpi:2023}
On the output side, the FPGA responds with Hrdata (0x12345678), matching the expected read data. The Paddr signal is output as 0x8C000000, representing the address sent to the APB bus. Similarly, the Pwdata value of 0x87654321 is forwarded as the write data to the APB peripheral. The Pselx[2:0] signal is asserted as 0101, indicating the active APB peripheral, while Hresp[1:0] is set to 0b10, signaling an error response condition. Additionally, the Pwrite signal is asserted as 1, confirming the APB write operation, and the Penable signal is activated as 1, enabling the peripheral transaction.

The Synthesis Area Report\cite{Mahesh_area_power:2023} for the BridgeTop design in Table ~\ref{tab:area_report} provides a detailed breakdown of resource utilization post-synthesis and place-and-route, highlighting the area contributions of combinational, sequential, and interconnect components. The design, synthesized using Design Compiler (DC) and implemented with IC Compiler II (ICC2), achieves optimized area usage while maintaining a balance between logic and routing resources.The BridgeTop module consists of 206 ports serving as input and output connections, enabling seamless interfacing with external components or systems. A total of 453 nets provide interconnections between logic elements, ensuring efficient signal flow across the design. The design incorporates 352 cells, which include both combinational and sequential elements.Out of the total cells, 114 combinational cells are dedicated to implementing purely logical operations such as AND, OR, and XOR gates. These cells perform critical combinational tasks without any memory elements. The design also\cite{Raj_area_fpga:2023} integrates 238 sequential cells, including flip-flops and registers, which serve to store and synchronize data with the clock signal, making them vital for maintaining state and timing consistency in the module.The 26 buffers/inverters included in the design help maintain signal integrity by amplifying signals over\cite{Sahoo_ahb_apb:2022} long interconnects and inverting signals where necessary. These components ensure reliable data propagation across the module.The Combinational area occupies 54.612001 units, reflecting the space utilized by purely logical gates. In contrast, the Noncombinational area contributes a larger footprint of 253.612809 units, highlighting the area occupied by sequential elements such as flip-flops and registers. Buffers and inverters consume an additional 6.482400 units, showcasing their\cite{Mehta_area_timing:2023} role in enhancing signal strength and stability.A significant portion of the design area, 477.019164 units, is allocated to net interconnects, emphasizing the complexity of routing signals between the logic components. Efficient routing strategies are critical to reducing delay and ensuring optimal design performance.The total cell area sums up to 308.224810 units, representing the combined space consumed by all active logic and sequential \cite{Narasimha_fpga_rpi:2023} components. The overall total area of the design, including interconnects and cell resources, amounts to 785.243974 units, reflecting the final physical footprint of the BridgeTop module.This area distribution highlights an efficient and balanced design, where interconnect routing and sequential components \cite{Sahoo_ahb_apb:2022}contribute significantly to the overall area. Such optimization is crucial for achieving scalability and integration in modern SoC architectures, particularly in applications requiring a mix of high-performance and low-power operations. This breakdown ensures that the BridgeTop module is area-efficient and ready for further implementation in advanced hardware systems.

Table ~\ref{tab:power_analysis_bridge} in the report presents the Power Analysis Report for the BridgeTop design, detailing the contributions of internal, switching, and leakage power across various power groups. The register group dominates the total power consumption, accounting for 97.70 of the total power with 334.8159 µW of internal power and a minimal switching power of 1.8795 µW. Leakage power within registers contributes 7.1027e+04 pW (approximately 71.027 nW), indicating the importance of optimizing register usage\cite{Mehta_area_timing:2023} to minimize power dissipation. The combinational group contributes a total of 7.9415 µW, which is 2.30 of the overall power, comprising 0.8345 µW of internal power, 7.0921 µW of switching power, and 1.4948e+04 pW of leakage power. Notably, power contributions from the iopad, memory, blackbox, clocknetwork, and sequential groups remain 0.00 µW, reflecting no activity or negligible power consumption for these components in the analyzed scenario. The combined switching power of 8.9716 µW highlights dynamic activity within combinational and register components, whereas leakage power across all components totals 8.5975e+04 pW (approximately 85.975 nW). The total power dissipation for the BridgeTop design is 344.7079 µW, with internal power being the dominant factor, followed by smaller contributions from switching and leakage power. This analysis highlights the significant role of registers and combinational logic in the design's overall energy profile.

The Timing Report for the BridgeTop Design (Table ~\ref{tab:timing_report}) provides a detailed breakdown of the signal propagation from the startpoint resetn (input port) to the endpoint FSM PRESENT STATE reg1 (a flip-flop). The timing path is driven by the clk clock source, with the clock network delay treated as ideal (0.00 ns). An input external delay of 0.10 ns is introduced for resetn, modeling the signal arrival relative to the clock edge. The path includes incremental delays contributed by various gates, starting with a buffer (U97/X) that adds 0.04 ns, followed by AND gates (U127/X, U130/X, and U102/X) with delays ranging from 0.03 ns to 0.04 ns, and an inverter (U202/X) contributing 0.03 ns. Two complex gates, U206/X (OAI22) and U207/X (AO221), add incremental delays of 0.02 ns each, bringing the total data arrival time to 0.30 ns.The clock period is 0.72 ns, and a clock uncertainty of -0.07 ns accounts for variations such as jitter or skew, reducing the available timing margin. Additionally, a library setup time of -0.01 ns is defined for the flip-flop. As a result, the data required time is calculated as 0.64 ns, with a final slack of 0.34 ns. The positive slack confirms that the design meets the timing constraints without violations, ensuring reliable operation within the defined clock cycle. This analysis highlights the incremental contributions of various components, the total delay accumulation, and the robustness of the design in achieving the required timing performance.

%}

\section{Conclusion}
\label{sec:conclusion}

% contribution

The implementation of the AHB to APB Bridge using the Artix-7 FPGA \cite{Shukla_spi_fpgapi:2022} and Raspberry Pi 4 Model B \cite{Narasimha_fpga_rpi:2023} successfully demonstrates seamless communication between high-speed AHB components and low-power APB peripherals. The system leverages SPI communication for data transfer and employs Verilog modules for protocol conversion, enabling efficient mapping of AHB signals to APB-compatible outputs. The Xilinx Vivado Design Suite facilitates design, behavioral simulation, and initial synthesis, ensuring functional correctness and optimization for FPGA hardware. Additionally, the Synopsys DC Compiler  is utilized for gate-level synthesis, providing detailed analysis and optimization of area, power,\cite{Joshi_icc2:2021} and timing. This step ensures that the design adheres to timing constraints while achieving resource-efficient implementation. Python scripts running on the Raspberry Pi manage the SPI interface for reliable data serialization, transmission, and validation, ensuring synchronization with the FPGA. Overall, the combined use of Vivado, Design Compiler, and SPI-based communication establishes a robust, scalable, and power-efficient solution for bridging high-performance AHB subsystems with low-power APB peripherals, showcasing its practicality and relevance in modern SoC designs.

\bibliographystyle{IEEEtran}
\bibliography{main}

%\vspace{10.5cm}

%\input{revision.tex}

\end{document}